\documentclass[twocolumn,showpacs,english,10pt,amsmath,amssymb,groupedadd
r e s s ]{revtex4}
\usepackage{graphicx}% Include Figure files
\usepackage{dcolumn}% Align table columns on decimal point
\usepackage{bm}% bold math
\begin{document}

\title{{\bf Molecular dynamics modeling of the effects of cementation 
on the acoustical properties of granular sedimentary rocks.}}

\author{Xavier Garc\'ia} 
\affiliation{Centro de F\'{\i}sica,
Instituto Venezolano de Investigaciones Cient\'{\i}ficas.  IVIC,
Apartado 21827,Caracas 1020 A,Venezuela. 
\\email:xavierbox@gmail.com
}
\author{Ernesto Medina}
\affiliation{
Department of Physics. Boston University,Boston, Massachusetts 02215.
}
\affiliation{Centro de F\'{\i}sica. Instituto Venezolano de
Investigaciones Cient\'{\i}ficas.  IVIC, Apartado 21827,Caracas
1020 A,Venezuela.
\\email ernesto@ivic.ve
}

%\date{\today}

\begin{abstract}

The incidence of cementation processes on the acoustical properties of sands is studied via molecular dynamics simulation techniques. In our simulations, we consider samples with different degrees of compaction and cementing materials with distinct elastic properties. The microstructure of cemented sands is taken into account while adding cement  at specific locations within the pores, such as grain-to-grain contacts.  Results show that acoustical properties of cemented sands are strongly dependent on the amount of cement, its relative stiffness with respect to the hosting medium, and its location within pores. Simulation results are in good correspondence available experimental data  and compare favourably with some theoretical predictions for the sound velocity   within a range of cement saturation, porosity and confining pressure. 
% With the proposed simulation techniques, the initial uncemented sand is built by simulating the settling process of sediments. Uncemented samples of different porosity are considered by simulating the mechanical  compaction of sediments due to overburden. Cementation is simulated through a particle based model which captures the underlying physics behind the phenomena.  

\end{abstract}

\maketitle
\section{Introduction}\label{introduction}
 After settling, mechanical and permeating  properties of sand 
 are modified by  processes  called 
 diagenesis.
%  \footnote{Here the term is used with a broad sense. In
%  some references,  diagenesis refers to processes that occur
%  after some depth in the stratigraphic column}. 
 Among other
 diagenetic processes, cementation is defined in a broad sense
 (Bryant, 1993) as the introduction of mineral phases   into  
 rock pores. Many mechanisms of cementation are discussed in
 the literature: minerals can be transported in suspension with
 underground waters and settle in the pores of a final hosting
 medium. In low flux zones, cements can form as a solid 
 precipitate on grain surfaces. Formation of clay bonds between 
 grains and grain interlocking as a product of the matrix 
 dissolution  (Gundersen et al., 2002) are among others, possible  
 cementation mechanisms.   
  
 Depending on their origin and  location within the pores, cements
 modify hosting rock properties in different manners.  It  has
 been experimentally observed  that, if cement is 
 deposited near grain-to-grain contacts, it can significantly  
 increase strength and stiffness of a granular material (Bernab\'e et al., 1992). In
 this case, grain rotation and displacement are  inhibited due
 to cement bonds and samples develops a frame that resists 
 compaction. In an experimental study with glass beads
 (Hezhu, 1992), the presence of epoxy cement at contacts  was
 observed to prevent  bead crushing during compaction. 
 Theoretically, this effect was attributed to a more uniform 
 stress distribution within contact areas due to cementation.
 This kind of cement is the  main responsible of rocks'
 cohesion. Therefore, the lack of this kind of cement is a
 potential handicap for the stability of boreholes. 

 A recent experimental study has addressed how
 cements can modify hosting rock microstructure and related
 hydraulic properties, such as permeability (Prasad, 2003).
 According to the authors, microstructure can be characterised
 through a geometric index $FZI=1/\sqrt{P_s}\tau S_v$, where $P_s$ is a pore shape
 related value, $\tau$ is the tortuosity and  $S_v$ the specific
 surface area per unit of grain volume. Such geometric index, 
 is then correlated with permeability and wave velocity. As an
 example, if cement is deposited uniformly on grains surfaces,
 specific surface area sharply increases. On the other hand, pore
 filling materials can also grow from grain surfaces towards pores,
 forming intertwined structures that increase 
 tortuosity (Manmath and Lake,  1995). Both mechanisms lead to different
 microstructures at the pore scale that affect rock hydraulic and
 mechanical properties.

 Despite the importance of the prediction of cement effects, this is often a
 difficult task since  the exact location of
 cement in pores is usually unknown. Although some insights can
 be obtained from microphotography or other high resolution
 methods, the usage of these techniques is limited when
 considering rocks at larger (formation) scales.  As an
 alternative, acoustical methods are of great value in the
 characterisation of the underground at larger extents, being 
 successfully used in the identification of underground
 structures and fractures. However, for the precise interpretation
 of acoustical logs it is necessary to advance in the understanding
 of the acoustics of porous media and identify the connections between  
 macroscopic acoustic observables, and microscopical
 petrophysical parameters. In this sense, due to the apparent
 linkage of rock stiffness and cementation, it seems plausible to
 use acoustic methods to estimate the type and amount of cement
 present in rocks. For this purpose, one must establish the
 relation between the acoustical observables, such as velocity and 
 attenuation,  and the cement saturation, its elastic properties and 
 localisation within pores among other parameters. 
  
 Experimentally, the acoustical effects of coating cement were 
 studied when considering artificially fabricated samples
 cemented with sodium silicate (Nakagawa and Myer, 2001). Samples showed
 increasing wave  velocities and attenuation with
 cement saturation. The acoustical effects of cement located near
 contacts, were studied experimentally by considering epoxy
 cement and glass beads (Hezhu, 1992). In this work, wave velocities sharply increased with cement saturation. Same results were reported 
 in (Dvorkin et al., 1994a; Dvorkin et al., 1994b) and references therein.

 From a theoretical point of view, Dvorkin et al.(1994a) 
 proposed a  model for cemented sands based on the effective medium
 approach. This model considers the elastic properties of a
 granular material whose porosity decreases by the addition of
 cementing  material near grain contacts. The model is based on the
 analytical solution of the related elasticity problem. 
 Theoretically, it is expected that small amounts of cement added 
 precisely at grain contacts, to sharply increase rock 
 stiffness.  Different cementation schemes, where cement is 
 added preferentially far from contacts, or as a coating on 
 grain surfaces were considered in (Dvorkin and Nur, 1996) and the results  
 summarised by Mavko et al.(1988). For reference, the main 
 formulae are also presented here in appendix \ref{emt}. This 
 theoretical formulation was proposed as an explanation for 
 some high-velocity high-porosity samples and 
 different velocity-porosity trends observed in  North Sea 
 sandstones (Dvorkin and Nur, 1996).

 Numerical  methods, such as molecular dynamics or finite element
 methods, represent a new approach to the study of complex
 multibody systems such as sand. In this approach, few numerical 
 studies consider the effects of cementation on the mechanical
 and hydraulic properties  of sands. In early papers, porous media
 was modeled using disks bonded by elastic   springs representing
 cement. The mechanical properties of this model
 were observed to depend on the bonding distribution and properties. 
 A three dimensional approach studied the brittle failure of
 cemented rocks under external load, modeling cement as breakable
 grain bonds  (Guodong et al., 2003).  
 
 Previous numerical studies considered cementation effects on
 permeability by emulating the microstructure of cemented rocks 
 and then solving flux equations in this geometry. Some
 approaches consider sand as a 
 pack of spherical particles whose radii is  extracted from a
 given distribution. The cementation process is  then simulated
 by increasing grains radii by a given amount, modeling the quartz cement overgrowth (Bryant et al., 1993; Bakke and Oren, 1997). More recent 
 studies account for  preferential growth of  cement to pores or
 throats by a more  elaborate algorithm that simulates cement
 overgrowth from grain boundaries to the surface of their Voronoi
 polyhedron (Schwartz et al., 1987). The resulting spatial correlations
 and other geometric characterisation was studied numerically by 
 Biswal et al.(1999).
 
 However, to the best of our knowledge, acoustics of sands is an
 open problem where numerical methods can be very useful 
 to  understand  the underlying physics behind the phenomena, to 
 test theoretical models and to improve their predictive
 capabilities. In our approach, we move a step further in
 the modeling  of acoustical response of cemented sands. We 
 propose a simple  particle based method (molecular dynamics)
 that considers the  amount of cement and its location within
 pores. 
 
 Particle  based methods or discrete methods differ from
 other well known  techniques such as finite elements in that no 
 continuous wave  equations are solved directly. Instead,
 molecular dynamics  mimics the underlying physics of wave
 propagation at the micro or meso scales using micromechanical
 interaction rules between discrete particles. This method
 involves solving  Newton's equations of motion for an N particulate 
 system  and global behaviour is obtained by the
 cooperative effect of interacting  particles (Rappaport, 1995).

 In our simulations we study the acoustical effects of soft and
 hard cements on a model sample. We consider the case when 
 cementing material is added preferentially to grain-to-grain 
 contacts, on grain surfaces or as a solid material  added 
 preferentially to pore  bodies. Our model sample mimics a well 
 sorted naturally occurring sand, and cementation is simulated 
 capturing the underlying physics governing the acoustic pulse 
 propagation on cemented sands.
         
 This paper is organised in three main parts: section 
 \ref{cementationprocess}, summarises the basic concepts related 
 with cementation  process. The following section \ref{procedures} 
 covers a  detailed description of the methods used in the simulations. Finally, simulation results are  presented and discused in section \ref{results}. Important  information regarding with the calibration of simulations and some theoretical 
 formulae are presented in appendix.

\section{Cementation Process}\label{cementationprocess}

 Clastic rocks are often cemented by addition of clay particles. 
These cements can be classified  in two broad categories:  
Detritical  and Diagenetical (Manmath and Lake,  1995). The former are 
transported in suspension with subsurface waters to a final 
location in a hosting medium where they deposit. This kind of 
cementing particles are mainly attracted to the sharpest zones of 
pore space, such as grain-grain contacts, where the surface 
forces are largest (Adamson and Gast, 1982).  These cements reinforce  
grain contacts, increases rock's stiffness and  sound 
propagation velocity.  Theoretically  is expected 
that even small amounts of soft cement added precisely at 
contacts, to sharply increase wave velocity propagation (Dvorkin et al., 1994b).

On the other hand, diagenetic cements form as precipitates of  solid material 
initially in solution in meteoric waters (or low mobility 
waters).  Due to its chemical origin, many varieties of these 
clays are possible, depending on the concentration of different 
species in pore fluids and chemical equilibrium at underground 
pressure and temperature. In this category, the so called pore 
lining clays deposit preferentially as coatings on grain 
surfaces, except at the grain to grain contacts (Guodong et al., 2003).  
The most frequently occurring pore lining cements are smectite, 
illite and chlorite. This kind of cements, if deposited 
homogeneously on grain  surfaces, leads to a sharp  increase in 
specific surface area of  pores while porosity reduces. 
Underground sections with this kind  of cementation can exhibit a 
decrease in permeability under the normal compaction trend 
(Prasad, 2003).   

Another possibility for clays is to grow as  long 
crystals from grain surfaces to pores. This kind of clays, such 
as radial illite and smectite,  are sometimes called pore 
bridging. The effect of these cements is mainly to reduce 
porosity and permeability while increasing tortuosity. As cement 
is mainly deposited away from contacts, it does not significantly 
contribute to sample stiffness until porosity is sufficiently reduced. 
 
Other cementing materials, such as quartz or calcite, are also 
present in rocks as cements. These cements can  locate 
preferentially at pore bodies (pore filling cements) forming tight 
structures. Quartz overgrowth is the formation of quartz 
crystals on the sand grains surface. Although crystals exhibit 
euhedral faces, the  overgrowth approximates also to concentric 
rims or coatings. 

\section{The Model}\label{procedures}
\subsection{Settling Process}\label{sedimentation}

The initial grain pack is built by following a ballistic 
algorithm to simulate the settling process of clastic sands 
(Bakke, 1997). The procedure models sand grains as spherical 
particles that fall on an initially flat surface from 
a fixed height in the $\hat{z}$ direction and random coordinates 
in the $xy$ plane. The radius of each particle is extracted from 
a given grain size distribution. In our simulations, the 
radius $R$ of each particle is selected with equal probability 
in the range $0.018 -0.02 cm$, modeling a well sorted 
sand.    
         
During settling, Newton's equations of motion are solved for 
grains falling under the influence of a gravity field $-g\hat{z}$, 
a viscous drag due to the aqueous medium and the  
interaction with other previously settled grains. 
At this stage, when two grains come into contact, 
they interact with a repulsive non-linear viscoelastic force 
${\bf F}_c$  given by: 
\begin{equation} 
\label{hertzforce} 
{\bf F}_c =\lbrace(\kappa _n \xi - \gamma_n{\dot\xi} \rbrace{\bf \hat
n}+{\bf F}_s.  
\end{equation} 

\begin{figure}[!h]
 \includegraphics[width=5.0cm]{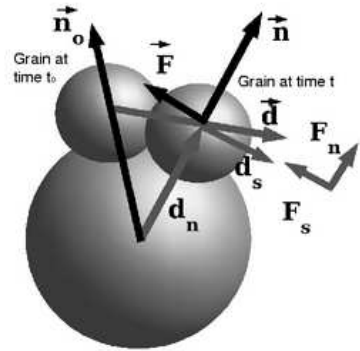}
 \caption[Schematic representation of contact forces]
 {\small\label{Fig1}
 Interaction forces are decomposed in normal and tangential 
 directions to the contact area.}
 \end{figure}

The first term in the Eq. \ref{hertzforce} represents the 
force  ${\bf F}_n$ normal to the contact area. For the labeled 
grains $1$ and $2$ with radii $R_{1}, R_{2}$ and positions ${\bf 
r}_1,{\bf r}_2$, $\xi={\rm max}(0, R_1+R_2-|{\bf r}_2-{\bf r}_1|)$ 
and   ${\bf \hat{n}}_{12}={\bf r}_1-{\bf r}_2 / |{\bf r}_1 -{\bf r}_2|$ is the unitary vector joining the grains centers  
(see Figure \ref{Fig1}).  
The normal stiffness of the contact $\kappa_n=4/3\sqrt{\xi R_{f}}E_{f}$ is computed according with Hertz theory, where $R_{f} = (R_1R_2)/(R_1+R_2)$ is an effective radius and 
$a_c=\sqrt{\xi R_f}$ is the radius of contact area. The effective Young's modulus $E_f$ is calculated  as $1/E_{f}=(1-\nu_1^2)/E_1 + (1-\nu_2^2)/E_2$, being $E_1$, $E_2$, $\nu_1$ and $\nu_2$ the Young modulus and Poisson's coefficient of the grains material (Love, 1944). 
The second term in braces in Eq. \ref{hertzforce} represents 
the the viscous forces for normal deformations, where  $\gamma_n= 
\tilde{\gamma}_n (\xi R_f)^{1/2}$ and $\tilde{\gamma}_n$ is a 
damping constant.

% \begin{figure}[!h]
%  \includegraphics[width=5.0cm]{xx.eps}
%  \caption[Schematic representation of contact forces]
%  {\small\label{Fig1}
%  }
%  \end{figure} 

%
\begin{equation} \label{shearforceeq} 
{\bf F}_s=-{\rm min}\left (|\kappa _s \zeta|,\mu
|F_n|\right)sign(\zeta){\bf \hat s}, 
\end{equation}
The shear force ${\bf F}_s$ in Eq. \ref{hertzforce}  
depends on contact history and cannot be entirely determined by 
the position of grains. Such force is given by Eq. 
\ref{shearforceeq}, where the tangential stiffness 
is $\kappa _s = 8a_c G/(2-\nu )$, shear modulus of the grains is given by $G$ and $\nu$  is the  Poisson's coefficient. The term $a_c$ denotes the radius of contact 
area previously defined and $\mu$ represents the Coulomb friction 
coefficient. The term  $\zeta$ denotes the component of the relative 
displacement vector  in the  tangential direction that took place since the 
time $t_0$, when the contact was  established. 
\begin{equation}\label{zeta}
\zeta(t)=\int_{to}^{t}v_s(t^{ \prime } )dt^{ \prime },
\end{equation}
where $v_s= {\bf v}_{ij} \cdot \hat{\bf s}$ and $\hat{\bf s}=\vec{\zeta}/|\vec{\zeta}|$. The vector ${\vec{\zeta}}$ can be 
computed from the displacement vector {\bf d} between the grains 
${\vec{\zeta}}={\bf d} - ({\bf d\cdot n}_0){\bf n}_0$  (see Figure \ref{Fig1}).

At this stage, periodic boundary conditions are imposed on 
gravity perpendicular directions to emulate an effectively larger 
system and the underground natural confinement of rocks in those directions.

The initial  pack models a high porosity $\phi=  
41.2\%$ sand, where contacts involve marginal contact areas. 
The size of the generated sample was $8 \times 8 \times 32$ in 
units of maximum grain diameter. Following previous results (Garc\'ia et al., 2004), this volume is several times larger than the minimum homogenisation volume for the porosity of a sample with the given  grain size distribution. 

Samples with different packing are obtained by simulating the naturally occurring compaction of sediments. Following Garc\'ia et al.(2004), this process is simulated by displacing towards the sample core, grains contained in frozen slabs at both ends of the sample in the $\hat z$ direction. Once such macroscopic strain in imposed, ample time is  allowed for  grains  to rearrange and relax accumulated stress.

\subsection{Cementation model}\label{cementation}
Once the initial sand pack is obtained, the cementation process is
simulated in several steps representing different stages in
the sand diagenetic history. At each step, a volume of cement is
added at specific locations  within pores. Cement is added in three different manners  that model cement with
distinct origins and their micro-geometric implications. 

The procedure followed to add cement starts by discrediting the three dimensional sample into cubic boxes of side length $c_w\approx R/10 $,
$R$ being  the average radius of matrix grains. The set of cells
intersects the sample voids and solid matrix, but just those
cells whose centre coincides with pore space are potential locations for  
cement. Cells are subdivided into three groups:

\begin{enumerate}
{\item \it Contact cells}  
are those that intersect two or more grains.  These, are located near 
grain contacts at the sharpest zones of pore space. 
{\item \it Surface cells} are those centred in pore space and contacting with 
at least with single grain. Contact cells are a subset of surface cells.  
{\item \it Body cells} are those centred in the pores without intersecting 
any grain.
\end{enumerate}

In practise, cement is added by marking  as cement saturated 
a fraction of pore centred cells.  This mark is 
achieved by locating a  cement particle at the centre of the 
cell. Each  cement particle represents a cubic 
subvolume in the sample where cement is allocated.

\subsection{Contact Cement}\label{contactcement}

To model the microgeometry associated with contact cement model, we propose an  algorithm of several steps that  
represent a different stage in the diagenetic history 
of the sample. Each step  involves porosity reduction due to the 
addition of cement to pores.  First, a small fraction of {\it contact 
cells} is marked as cement saturated. During the following steps, an 
increasing number of {\it contact cells} are filled with cement until 
no  {\it contact cells} are left. 

In practise, a cell are marked as cement saturated by adding a single 
cement particle to the centre of the cell. As this particle represents 
a cubic subvolume $c_w^3$ of cement within the sample, we  assign to it 
a mass $m_c=\rho _c c_w^3$, where $\rho _c$ is the volumetric density of 
the cementing material.
  
After all {\it contact cells} are filled, further cementation steps 
involves the addition of cement to selected cells, neighboring those 
previously filled. First, we list all empty neighboring cells of those 
already filled with cement. 
Then, among these empty cells we select those lying  closer to the grain matrix, and fill them with a new cement particle. Thus, when all {\it contact cells} are filled, this procedure promotes the cement accumulation from contacts to {\it surface cells}. 

\begin{figure}[!h]
\includegraphics[width=7.0cm]{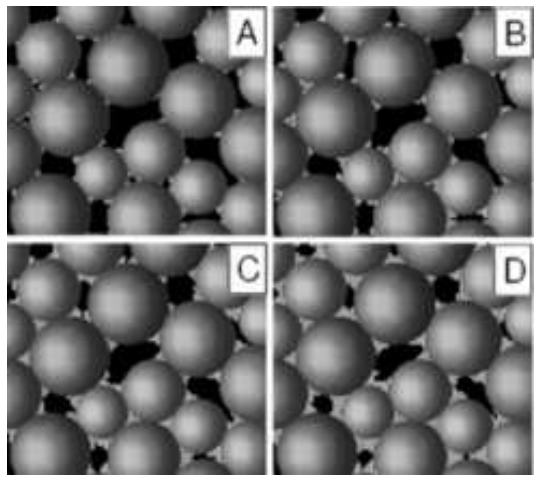}
\caption[Contact cement deposition]
{\small\label{Fig2} 
Figures $A$,$B$: an increasing number of {\it contact cells} is 
filled with cement until no {\it contact cells} left. Then, cement grows from  
contacts to surfaces in Figures $C$ to $D$. {\bf In the Figure, $2D$ samples are shown for 
clarity. Nevertheless, results reported here refer to the three-dimensional samples.}
}
\end{figure}

A sequence of steps where cement was added according to the described procedure is shown in Figure \ref{Fig2}. the Figure shows a two dimensional example for illustrative purposes. However, results reported here refer to three dimensional samples and smaller cells (cells width $c_w \approx$ $1$ $th$  of average grain radius).

% \begin{figure}[!h]
% \includegraphics[height=6.0cm]{xx.eps}
% \caption[Contact cement deposition]
% {\small\label{Fig2} 
% }
% \end{figure}

\subsection{Coating cement}\label{coatingcement}

\begin{figure}[!h]
\includegraphics[width=7.5cm]{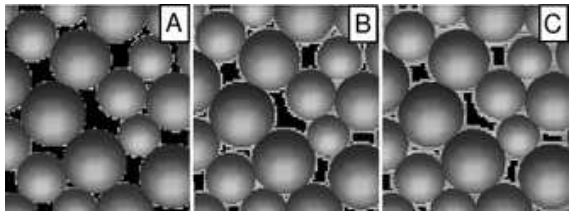}
\caption[Sequence of steps for coating cement]
{\small\label{Fig3} 
From step $a$ to $c$ cement is  added to grain surfaces as a 
coating. This procedure simulates the homogeneous precipitation 
of solid material on the surface of pores. 
{\bf Figure shows a two dimensional case for clarity, reported 
results are for three dimensions though.}
}
\end{figure}

To model the  microstructure related with coating cement, we  mark each cell  with an index proportional to its distance to the solid  surface. At each cementation step, all empty  cells are checked and those with the  smallest index (closest to  matrix surface) are filled with a cement particle. In the first  steps, cementing material is added on grain surfaces. As is shown in  Figure \ref{Fig3}, further cementation steps add cement deeper into the  pore bodies so that rims appear around grain surfaces. 

% \begin{figure}[!h]
% \includegraphics[height=5cm]{xx.eps}
% \caption[Sequence of steps for coating cement]
% {\small\label{Fig3} 
% }
% \end{figure}

The successive cementation steps simulate several stages of the
cementation history of the sample, where a change in the
chemical equilibrium conditions causes precipitates to deposit 
uniformly on the grain surface.

\subsection{Pore body cement}\label{friable}

Simulation of pore body cements  begins with the addition of cement particles to a fraction of surface cells selected at 
random. In the following steps, we check the empty cells contacting those already filled with cement. Each empty cell is then  
filled with a cement particle according to a probability
$p$, proportional to the cell distance to grain surface. This 
procedure, simulates a preferential  grow of cement from grains surface to pore bodies.  

\begin{figure}[!h]
\includegraphics[width=7.5cm]{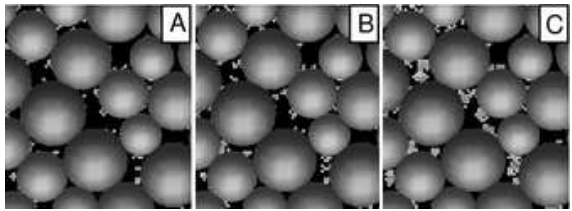}
\caption[Sequence of steps for friable cement]
{\small\label{Fig4} 
Figure $a$: initial seed of cement. Figures $b,c$: further 
cementation steps when cement is added preferentially 
at pore bodies.
{\bf Figure shows a two dimensional cases for clarity. However, results refer to three dimensional samples.}}
\end{figure}

As shown in Figure \ref{Fig4}, the initial cemented
cells behave as seeds for clusters that grow from the grain 
surface to pore bodies. The described procedure guarantees the
continuity of the cement phase and simulates cementing materials in
the form of long crystals attached to grain surfaces. 

%  \begin{figure}[!h]
%  \includegraphics[height=4cm]{xx.eps}
%  \caption[Sequence of steps for friable cement]
%  {\small\label{Fig4} 
% }
% \end{figure}

Cell width  defines the resolution limit of 
the simulations. If cells are too wide, poor resolution is 
obtained and porosity decreases sharply to zero after a few 
cementation steps. Small cells allow more control of the final 
 porosity of the sample at expenses of an increase in the  
 computational cost. More details about the effect of resolution  
 on the computational cost are found in the appendix 
 \ref{resolutioncost}.

\subsection{Interaction}\label{interaction}

After a given porosity is reached due to cementation, we 
simulate the acoustical pulse transmission through the sample 
using Molecular Dynamics techniques. As Molecular Dynamics involves 
the explicit solution of Newton's  equations of motion for a particle 
based system, we need to consider the interaction forces between 
the particles of the different constituents of the aggregate. In our 
model, particles conform the quartz matrix and cement. We do not 
consider fluid  particles that may be in pore volume.
% In this sense, the elasticity of the sample 
%in  governed by solid phase. 
This approximation is justified if 
we suppose the fluid saturating pores to be air free to drain while sample 
strains. The effect of a more incompressible saturating fluid such as water 
or oil could be included and it is proposed for  future work. In this 
way, our simulation techniques allow us to isolate the acoustical 
effect of cement from other agents that  modify the acoustics of 
a real sample. With this approximation, we need only  consider three kinds 
of interaction forces between particles: grain to 
grain forces, interaction between cement particles,  and interaction forces  
between cement particles and grains.

In our model, the addition of cementing material  is assumed not to  disrupt the internal stress state of the sample. This assumption is justified since cementation is a relatively slow process when compared to compaction (Bernab\'e et al., 1992). 
Therefore, cement is added to pores at mechanically stable locations 
after grain-to-grain contacts are formed. 
When a perturbation of this equilibrium state occurs, changes in 
the internal stress field of the sample are produced. At the pore 
scale, grains change their relative positions and  interparticle forces
change by an amount $\Delta{\bf F}_c$ with respect to the initial
equilibrium network. 

If after a perturbation a pair of contacting grains labeled $1$ and $2$ 
at  positions ${\bf r}_1$ and ${\bf r}_2$, displace to 
${\bf r}_1+{\bf dr}_1$ and ${\bf r}_2+{\bf dr}_2$ respectively, 
their relative displacement is calculated as  ${\bf d}={\bf dr}_2-{\bf dr}_1$. 
The component of the displacement along the normal direction of the contact 
is given by $\Delta\xi={\bf d}\cdot {\bf \hat{n}},$ being ${\bf \hat{n}}=({\bf r}_2 -{\bf r}_1)/\vert{\bf r}_2 -{\bf r}_1\vert$ the unit vector joining  grains centre at 
equilibrium. The tangential displacement is calculated as $\vec{\Delta\zeta}= {\bf d}-({\bf d}\cdot{\bf \hat{n}}){\bf \hat{n}}$. This vector defines the tangential direction 
${\bf \hat{s}}=\vec{\Delta\zeta}/\Delta\zeta$ 
with $\Delta\zeta=\vert\Delta\vec{\zeta}\vert$.
During this stage, the force change 
${\bf \Delta F}_c$ due to relative displacement of contacting grains is computed 
according to:

\begin{equation}\label{linearhertz}
{\bf \Delta F}_c = \lbrace(k_n  \Delta\xi  
 - \gamma_n \dot{\Delta\xi} )\rbrace {\bf \hat n}+{\bf \Delta
F}_s.
\end{equation} 
Term $\kappa_n$  
%=4/3 \sqrt{ \xi _0 R _f}E_f$ 
in Eq. \ref{linearhertz} is the elastic stiffness for normal deformations, defined 
previously in equation \ref{hertzforce}. However, in this stage we assume a negligible change 
in the contact area due to the acoustic pulse.   
%  $\xi_0=max(0,R_1+R_2-\vert {\bf r}_1-{\bf r}_2)$ is the initial overlapping of contacting particles and  $\sqrt {\xi_0 R_f}$ is the radius of the initial contact area according  with Hertz theory. 
 This approximation constitutes  a linearization of Hertz force, that  
 represents the normal 
grain$-$grain elastic interaction that acts on contact. The second term in braces in Eq. \ref{linearhertz}, describes 
the viscous forces for normal deformations. In our simulations, we have adjusted  this parameter to obtain a normal restitution coefficient $e_n=0.9$, in
agreement with experimental results (Kuwabara and Kono, 1987; Schafer et al., 1996).

The term ${\bf\Delta F}_s$ in Eq. \ref{linearhertz} represents the change  in 
shear force. To compute this term, we  
assume  that no sliding between grains occurs during pulse 
transmission. In such case, shear force is considered 
viscoelastic and calculated according  to 
\begin{equation} \label{linearshearforce} 
{\bf \Delta F}_s= \lbrace k_s  \Delta\zeta  
 - \gamma_s{\dot{\Delta\zeta}} \rbrace {\bf \hat s}.
\end{equation}
The first term in the Eq. \ref{linearshearforce} represents
the elastic component of the shear force.
% , $\Delta\zeta$ being  
% the relative displacement of particles from their equilibrium 
% positions in a direction ${\bf \hat{s}}=\vec{\Delta\zeta}/|\vec{\Delta\zeta}|$  perpendicular 
% to ${\bf \hat{n}} $. 
%
As no sliding is assumed, the 
tangential stiffness is computed according with Hertz-Mindlin
theory $\kappa _s = 4a_cG /(2-\nu )$. The term $G$ denotes the
grains' shear modulus, $\nu$ is the material Poisson's ratio and
$a_c$ is the radius of the initial contact area 
defined previously. The second term in the Eq. \ref{linearshearforce} represents 
the viscous forces for shear deformations. Parameter $\gamma_s$ was adjusted to 
obtain a tangential restitution coefficient $e_s=0.9$. 
The parameters to compute forces, were chosen to model quartz 
sand grains (see Table \ref{quartzconstants}).  
%

% \vspace{2cm}
% \begin{center}Table I\end{center}
% \vspace{1cm}
{\Large
\thispagestyle{empty}
\begin{table}[tbp]
\caption{Parameters used in Eq. \ref{linearhertz} and
\ref{linearshearforce} to compute forces between quartz 
grains (Mavko et al., 1988).}
\label{quartzconstants}
\renewcommand{\arraystretch}{2.0}

\begin{tabular}{l  c  c  c c c c }
\tableline 
  Parameter    &   Symbol &   Value \\
\tableline 
  Density$(gr/cm^3)$         &  $\rho_g$  &  2.65 \\
  Shear modulus(GPa)         &  $G$       &  44 \\
  Poisson coefficient        &  $\nu$     &  0.08 &\\
  Friction coefficient       &  $\mu$     &  0.3 &\\
\tableline
\end{tabular}
\end{table}

}

To compute forces between cement particles, we follow a procedure 
described by O'Brien and Bean (2004). In this procedure, cement particles are considered as nodes on the cubic lattice that discretise the sample. 
When a cement particle  displaces from its initial position on the 
lattice, it interacts with its $18$ closest neighbours in the lattice as 
shown in Figure \ref{Fig5}. This interaction force is calculated according 
to:
\begin{figure}[!h]
\includegraphics[width=6.0cm]{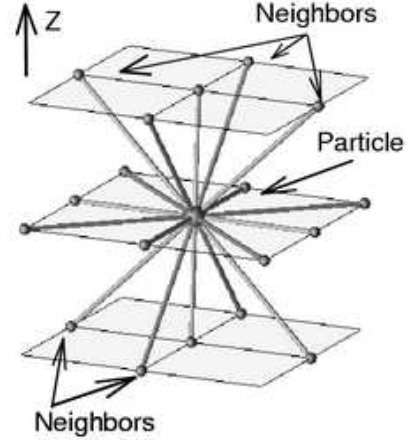}
\caption[Lattice representation of cement and interacting pairs of particles]
{\small\label{Fig5}
Each cement particle interacts with cement particles present in 
each of its closest eighteen nodes. Interaction force is 
decomposed in a component that acts  along the line joining 
cells centre and a bending term.
}
\end{figure}

% \begin{figure}[!h]
%  \includegraphics[width=7.0cm]{xx.eps}
%  \caption[Lattice representation of cement and interacting pairs of particles]
%  {\small\label{Fig5}}
% \end{figure}

\begin{equation}\label{cementforces}
{\bf f_c}= C_n \left[{\bf u}_{12} . {\bf\hat{n}}\right ] {\bf
\hat{n}}
 + \frac{C}{|{\bf x}_{12}|^2} {\bf u}_{12}, 
\end{equation}
where  ${\bf x}_{12}$ 
denotes the vector joining particles $1$ and $2$ on the undisturbed lattice 
and ${\bf \hat{n}}={\bf x}_{12}/|{\bf x}_{12}|$. The vector ${\bf u}_{12}$ denotes the relative displacement of particles from their relative positions on the 
undisturbed lattice. The first term is the normal force with elastic constant 
$C_n$. The second term in Eq. \ref{cementforces} is the bending term  and
$C$ is an elastic constant for the interaction (O'Brien and Bean, 2004). The force 
given  in Eq. \ref{cementforces} can be expressed in a form similar  to Eq. 
\ref{linearhertz}. However, we  decided to remain as close as 
possible to the original reference for clarity.

Following O'Brien and Bean (2004), energy of the elastic lattice can be
compared with the elastic energy of an elastic continuous.
The shear elastic modulus $G_c$ and Lam\'e constant $\lambda_c$ of the lattice 
can be expressed in terms of $C_n$ and $C$ according 
with the relation:

\begin{equation}\label{cemparams}
\begin{array}{l}
\lambda_c = \frac{C_n}{c_w} - \frac{2C}{c_w^3},\\
\\
G_c = \frac{C_n}{c_w} + \frac{2C}{c_w^3},
\end{array}
\end{equation}
where $c_w$ is the lattice parameter (in our simulations 
$c_w\approx 1$ $th$ of average grain radius).
 
\begin{equation}\label{cemvels}
 \begin{array}{l}
 V_{pc}^2 = \frac{1}{\rho_c}
 \left(\frac{3C_n}{c_w} + \frac{2C}{c_w^3}\right), \\
 \\
 V_{sc}^2 = \frac{1}{\rho_c}\left(\frac{C_n}{c_w} +
 \frac{2C}{c_w^3}\right).\\
 \end{array}
\end{equation}

From linear elasticity, it is possible to express the wave velocities 
in the cement phase as in Eq. \ref{cemvels}, where $V_{pc}$ and $V_{sc}$ 
are respectively the $P-$wave and  $S-$wave velocity of the cementing 
material. The term $\rho_c$ 
is the volumetric mass density of cementing material. As each 
cement particle represents a cubic subvolume $c_w^3$ of 
cement, such density is related to the mass of cement particles $m_c$.
\begin{equation}\label{cementparams2}
\begin{array}{l}
 C_n = \frac{1}{2}\rho_c\left({V_{pc}}^2 - {V_{sc}}^2 \right )c_w\\
 \\
 C =  \frac{1}{4}\rho_c\left( 3{V_{pc}}^2 - {V_{sc}}^2 \right) {c_w}^3\\
 \\
 m_c=\rho_c c_w^3.\\
 \end{array}
\end{equation}

For a cement of given mass density $\rho_c$ and wave velocities $V_{pc},V_{sc}$, 
the mass of cement particles $m_c$ and elastic constants in Eq. 
\ref{cementforces} are determined through Eq. \ref{cementparams2}. In our simulations, we considered the hard cement and  
soft cement limits compared to quartz. Parameters used in both cases are shown in Table
\ref{cementconstants}. The hard cement case, models a quartz-like cement 
such as that formed by quartz dissolution of matrix grains. Soft cement models a material added to pores such as soft clays. To simplify, we considered the same volumetric mass density for cement in both cases.

%
% \vspace{1.5cm}
% \begin{center}Table II\end{center}
% \vspace{1cm}
%
%
\begin{table}[tbp]
\caption{Elastic parameters and mass density used to model different cementing materials.}
\label{cementconstants}\renewcommand{\arraystretch}{2.0}
\begin{tabular}{ l c c c c c }
\tableline 
    Parameter &  Symbol &  Hard cement &  Soft cement \\
\tableline 
  P-Velocity$(Km/s)$    &  $V_{pc}$  &  3.0   &  1.5   \\
  S-Velocity$(Km/s)$    &  $V_{sc}$  &1  .8   &  0.9   \\
  Density$(gr/cm^3)$	   &  $\rho_c$  &  2.65  &  2.65    \\
  Poisson Coef.         &  $\nu_c $  &  0.22  &  0.22 \\
\tableline
\end{tabular}
\end{table}

The third kind of interaction force arises when  grains
displace relative to cement  particles contacting their surface.
This interaction in also computed  by Eq.\ref{linearhertz}, 
but the elastic constants $\kappa_n$ and $\kappa_s$ are now 
computed considering the Young's modulus and Poisson coefficient of 
quartz and cement. The diameter of contact area is 
taken as the cement cell width $c_w$. In this manner, a grain feels 
a force due its contacting grains and due to cement particles 
contacting its surface.

\subsection{Acoustics}\label{acoustics}

\begin{figure}[!h]
\includegraphics[width=6.0cm]{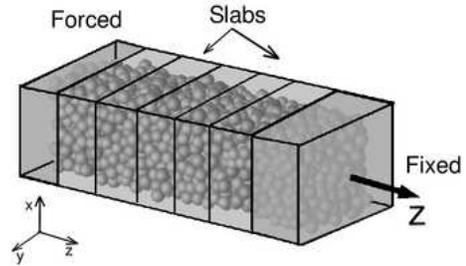}
\caption[Sample is subdivide in parallel slabs where stress
averages are calculated as function of time. One ending slab is
forced while the opposite is frozen.]
{\small\label{Fig6}
Sample volume is discretised into parallel slabs along $\hat{z}$
axis. At one end, a square wave displacement is imposed while the opposite slab is frozen. Average forces on grains
contained in each slab are calculated as function of time to
track pulse position and  amplitude.}
\end{figure}

In order to perform acoustic tests, the sample is subdivided along
the propagation axis $\hat{z}$ into parallel slabs of $2$ largest 
grains diameters. A square wave displacement is imposed on 
grains contained in one end slab of the sample while the other ending slab is kept fixed as 
is shown in Figure \ref{Fig6}. 
The former boundary condition, emulates the setup of an acoustic measurement. The fixed boundary condition on the opposite end  prevents the sample from displacing  as a whole.
 
%  \begin{figure}[!h]
%  \includegraphics[width=7.0cm]{xx.eps}
%  \caption[Sample is subdivided in parallel slabs where stress
%  averages are calculated as function of time. One ending slab 
%  is forced while the opposite is frozen.]
% {\small\label{Fig6}
% }
% \end{figure}

The imposed strain displaces grains from their original
equilibrium positions and disrupts the initial force network. This
perturbation propagates in the form of an acoustic pulse with a well
defined central frequency. To keep track of the pulse position and
amplitude, we monitor changes in the  forces felt by grains as
function of time and position. The average of the force along $z$
over the grains contained in each slab at time $t$ is taken as 
the acoustical signal  $A(z,t)$ at the position of the slab centre
$z$. This value is calculated according with the relation: 
\begin{equation}\label{signal}
A(z,t)=\frac{1}{N_i}\sum_i^N {f_i}, 
\end{equation}
where term ${f_i}$ is the force change at time $t$ on the $ith$ grain or cement
particle in the slab. The summation index runs over  the $N$
particles in the slab.
This procedure simulates a set of acoustical detectors distributed 
along the sample. 

\begin{figure}[!h]
\includegraphics[width=7.0cm]{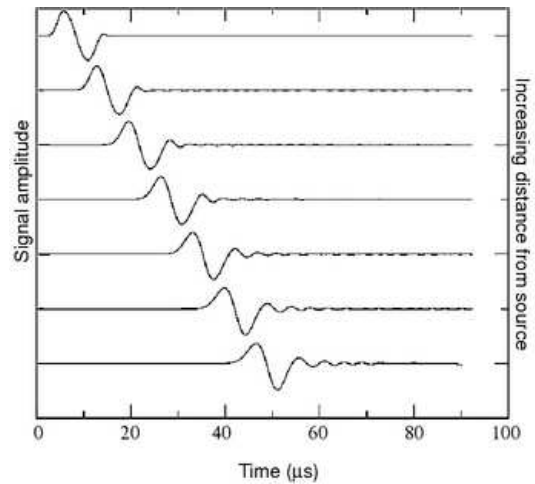}
\caption[Acoustical signals received in simulated detectors.]
{\small\label{Fig7}
 Signals recorded, at the simulated acoustical
 detectors, correspond to the average force along $\hat{z}$ of
 grains and cement particles  in different slabs of the sample.
 From top to  bottom, signals are recorded at detectors located
 increasingly farther from the source.
}
\end{figure}

As Figure \ref{Fig7} shows, the 
position in time of the stress pulse can be tracked by comparing
signals received in different detectors. 

% \begin{figure}[!h]
% \includegraphics[width=7.0cm]{xx.eps}
% \caption[Acoustical signals received in simulated detectors.]
% {\small\label{Fig7}
% }
% \end{figure} 

\begin{equation}\label{velocity}
V_p=\frac{L}{T_2-T_1} +- \frac{L\Delta T + T\Delta
L}{(T_2-T_1)^2},
\end{equation}

 Mean pulse velocity $V_p$ is calculated according to 
 Eq. \ref{velocity}, by comparing the arrival times  
 $T_1$ and $T_2$ of the signal maximum in two detectors separated by a distance of $L$.
The error in  pulse position $\Delta L\approx R$ is  estimated as the half width of slabs $\approx 2R$.  Error in time arrival  $\Delta T$, is taken as the  time spacing between two  
consecutive calculations of force average in the slabs. 

\section{Results}\label{results}

We simulate the acoustic pulse propagation through a cemented sand. 
Cement is preferentially added near grain-to-grain contacts following the procedure described in section \ref{contactcement} and its elastic properties are given in the column labeled as {\it Hard Cement} in Table \ref{cementconstants}. These parameters are suited to model a quartz-like cement and common used cements in experiments.  This initial system mimics a well sorted and non compacted sand with porosity $\phi= 41.2\%$ and point grain-to-grain contacts. We refer to this system as the non-compacted sample. Porosity reduction of the initial high porosity sample is achieved in several steps in which varying amounts of cement are 
added within pores. Acoustic pulse propagation is simulated for different degrees of cement saturation $S_c$.

\begin{figure}[!h]
\begin{center}
\center{\par\centering
\includegraphics[width=8cm]{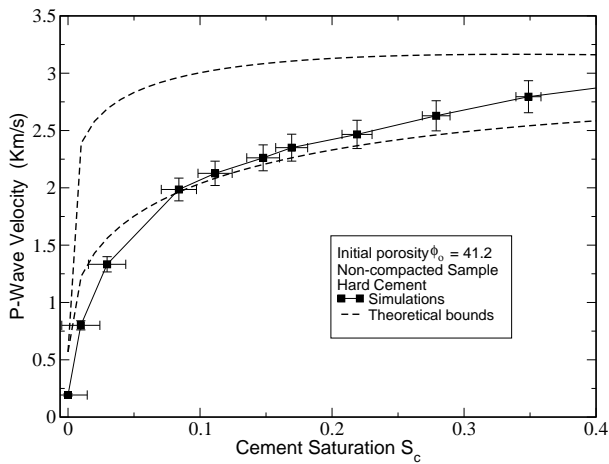} \par}
\caption[P-wave velocity versus porosity computed for 
different cement saturations $S_c$ of cement added near contacts in an initially uncompacted sample.]
{\small\label{Fig8}
Small amounts of contact cement contributes to increase stiffness of the composite. As more cement is added, it locates farther from 
contacts and stiffness increases at an smaller ratio. Consequently, $V_p(S_c)$ trend shows a characteristic downward concavity.  
}
\end{center}
\end{figure}

%  \begin{figure}[!h]
%  \center{\par\centering
%  \includegraphics[width=7cm]{xx.eps} \par}
%  \caption[P-wave velocity versus porosity computed for 
%  different cement saturations $S_c$ of cement added near contacts in an initially uncompacted sample.]
%  {\small\label{Fig8}
%  }
%  \end{figure}

As shown in Figure \ref{Fig8}, we observe an increase in sound velocity when an  small amount of cement is added preferentially 
near grain contacts. This trend occurs because cement increases the restoring force of contacts and bridges  grains initially separated. As a result, stiffness of the sample increases at a higher rate  than porosity is reduced. Consequently, $V_p$ trend shows a downward concavity. As has been  pointed  before by Dvorkin and Nur (1996), this is a consistent explanation to some observed high porosity-velocity samples.

When  more cement is added, porosity reduces and particles tend to locate progressively 
farther from contacts. In this case, stiffness increases at an slower rate than in the higher  porosity range, when cement particles locate 
precisely at contacts. The net result is a decreasing slope of the velocity-porosity curve 
as porosity is reduced. 

To compare our results with effective medium approximations  (Dvorkin et al., 1994a; Dvorkin and Nur, 1996), we plot in Figure \ref{Fig8} the theoretical bounds for the sound velocity of the cemented sample. Both theoretical curves were obtained with the formulae given in 
appendix \ref{emt} for the simulation parameters. The upper 
trend (high velocity) corresponds to the idealised case when all 
cement is precisely added at contacts. The low velocity trend, 
corresponds to the scheme in which cement is added uniformly on the grain surface. 

As shown in the Figure \ref{Fig8}, obtained velocity-porosity trends for the non-compacted sample with contact cement, have the same  qualitative behaviour than theoretical  predictions. However, theory predicts a sharp velocity rise when small amount of cement is first added. Beyond the initial velocity rise, theoretical curves tends to saturate. Instead, we observe a soft continuous velocity increase with  cement saturation $S_c$.

For a set of parameters close to ours, previous experimental studies  reported  a similar qualitative behaviour for a system of beads cemented with solidified epoxy (Hezhu, 1992; Dvorkin et al., 1994a), but  the experimental results lay between both theoretical curves (Dvorkin et al., 1994a). 
We attribute this discrepancy with our results, to the  stiffness of the initial uncemented sample. Theoretically, a finite non vanishing velocity is expected for the uncemented samples regardless of the effects confining
 pressure. For the case shown in the Figure \ref{Fig8}, the minimal theoretical velocity is higher than that obtained in our simulations.  Experimentally, results in the aforementioned experiments deal with compacted samples, that uncemented, propagate sound at velocities higher than the minimum predicted in 
the theory (Hezhu, 1992). On the other hand, in our settling algorithm  grains settle due gravitational forces and no confining pressure is applied. 
As a result, we obtain very loose samples with marginal contact  areas where P-Wave velocity can be quite smaller. 
% ($V_{p}\approx0.2 Km/s$). 
%

With the aim to better reproduce common experimental setups used for rocks, we simulated the compaction of the uncemented sample to a porosity $\phi=38.6\%$. At this porosity, confining pressure was recorded as $P_c\approx 30MPa$ and sound velocity was observed to increase from  $V_{p}=0.2 Km/s$ to $V_{p}=1.7Km/s$, similar to the minimal velocity reported  in the aforementioned experiments (Dvorkin et al., 1994a). Then, the cementation process was simulated as in the previous case and acoustic tests were performed for different cement saturations. 

\begin{figure}[!b]
\begin{center}
\center{\par\centering
\includegraphics[width=8cm]{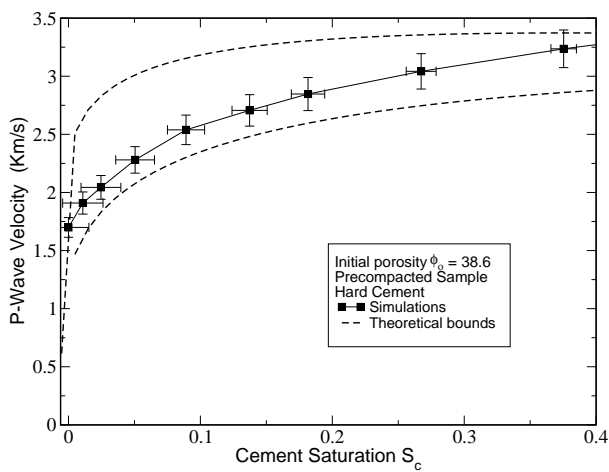} \par}
\caption[P-wave velocity versus porosity computed for 
different cement saturations in a precompacted sample of porosity $\phi=38.4$.]
{\small\label{Fig9}
Pulse velocity for different cement saturations. Uncemented sample is compacted with $30MPa$ confining pressure. After compaction, cement is added preferentially near contacts.  Theoretical predictions can be good estimates of velocity bounds for precompacted samples.
}
\end{center}
\end{figure}

%  \begin{figure}[!h]
%  \center{\par\centering
%  \includegraphics[width=7cm]{xx.eps} \par}
%  \caption[P-wave velocity versus porosity computed for 
%  different cement saturations in a precompacted sample of porosity $\phi=38.4$.]
%  {\small\label{Fig9}
% }
% \end{figure}

As expected, Figure \ref{Fig9} shows that for zero cement saturation, velocity is higher in the compacted sample than in 
the uncompacted sample. 
Additionally, the Figure shows that in the case of the compacted sample, velocity increases at a slower rate when cement saturation $S_c$ increases. The net result is that the velocity trend for the compacted sample lies between theoretical 
curves.  In this case, for the range of cement saturation $S_c$ 
shown in the Figure \ref{Fig9}, obtained results are in good quantitative agreement with available experimental data reported 
in previous works (Hezhu, 1994; Dvorkin et al., 1994; Dvorkin and Nur 1996). 
It should be noted that for the case of the compacted sample, theoretical curves are slightly different from those of the uncompacted sample, since coordination and porosity are different. 
Nevertheless, these  variations are relatively small. 

These results suggest that confining pressure of the initial uncemented sample is a crucial parameter which determines the velocity trends obtained after adding small amounts of cement. 
This is an intuitive result since for loose sediments close to suspension, stiffness can be negligible. Then, any stiffness 
increase due to cementation can be several times the initial 
one and represent a large fraction of the final stiffness of 
the cemented sample. When samples are previously compacted, the initial hosting medium for the cement becomes stiffer (despite fracturing). In this case, further addition of small amounts 
of cement can represent an smaller fraction of the final stiffness of the sample. 

Results in Figure \ref{Fig8} and Figure \ref{Fig9}, suggest that contact cement theory may  work well for samples which are cemented after some compaction is achieved. However, it may overestimate sound velocity for the case of cement added to weakly compacted media. 
The latter case, can be that of artificially cemented granular materials such as young  sediments cemented with perforation mud,  or laboratory produced materials. We believe that the  case where cement is added after some compaction is  the most realistic for rocks,  since chemical cementation is achieved after some pressure-temperature conditions are reached with burial. 
\begin{figure}[!h]
\includegraphics[width=8.0cm]{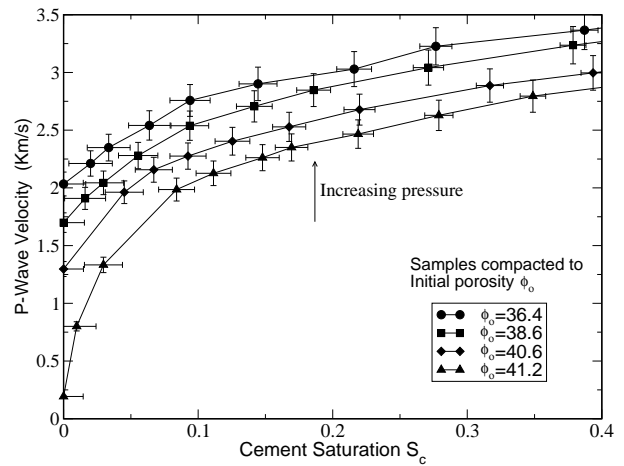}
\caption[P-wave velocity versus cement saturation in a set of samples precompacted to different initial porosity $\phi_o$.]
{\small\label{Fig10} 
Each curve represents a possible diagenetic path for a granular material. 
Samples are first compacted to a porosity $\phi_o$ due to burial. Then,  cementation process begin. The more compacted the samples, the smaller the relative increase of velocity with $S_c$.
}
\end{figure}

% \begin{figure}[!h]
% \includegraphics[width=7.0cm]{xx.eps}
% \caption[P-wave velocity versus cement saturation in a set of samples precompacted to different initial porosity $\phi_o$.]
% {\small\label{Fig10} 
% }
% \end{figure}

%
This behaviour is further illustrated in Figure \ref{Fig10} where    results are shown for several cases in which cementation starts after some compaction. In this Figure, each trend may represent a  possible diagenetic path followed by a granular material. The starting framework in all cases is  that of a very loose material of porosity  $\phi=41.2\%$, representing sediments after settling.  The initial loose pack is then compacted to a porosity $\phi_o$, simulating the mechanical effects of burial. After compaction, a relatively hard cementing material is added in varying amounts. 
\begin{figure}[!h]
\begin{center}
\center{\par\centering
\includegraphics[width=8cm]{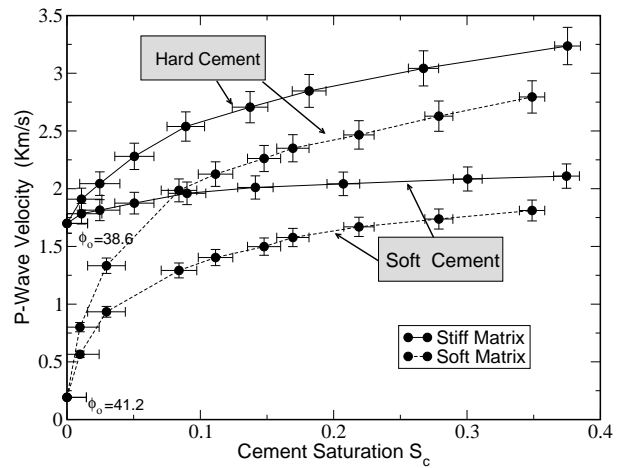} \par}
\caption[P-wave velocity versus porosity for the case of cement added at grains contacts of an uncompacted sample. Figure shows the case of  soft and hard cement]
{\small\label{Fig11}
Velocity increase in soft samples depends mainly in the amount of cement. For stiff samples, velocity increase depends strongly on the elastic properties of cement.}
\end{center}
\end{figure}

% \begin{figure}[!h]
% \center{\par\centering
% \includegraphics[width=7cm]{xx.eps} \par}
% \caption[P-wave velocity versus porosity for the case of cement added at grains contacts of an uncompacted sample. Figure shows the case of  soft and hard cement]
% {\small\label{Fig11}}
% \end{figure}

As shown in Figure \ref{Fig11}, we observed the same qualitative behaviour when considering cementing materials with different elastic properties. In the Figure, sound velocity in the uncompacted sample increases by a factor of $\approx 10$ when a relatively hard cement saturation varies from $0-0.1$. For the same cement saturation, a relatively soft cement increases sound velocity in a factor of $\approx 7$. In the case of a compacted (stiffer) sample, this variation approximates to $\approx 25\%$ in the case of the hard cement and $\approx 12\%$ for the soft cement. According with these results, velocity rise in soft samples mainly depends on the amount of cement while in stiff samples the most relevant factor is the type of cement.

\begin{figure}[!h]
\begin{center}
\center{\par\centering
\includegraphics[width=8cm]{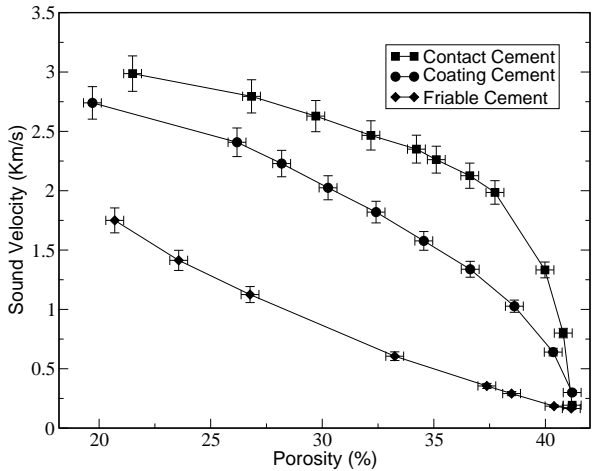} \par}
\caption[Acoustic response for three different cementation schemes.]
{\small\label{Fig12}
Acoustical response depends on microstructure details. Contact  cement leads to a downward concavity in $V_p(S_c)$ trend with decreasing slope as cement saturation increases. Friable cement leads to the opposite behaviour. Coating cement corresponds to an intermediate case. 
}
\end{center}
\end{figure}

% \begin{figure}[!h]
% \center{\par\centering
% \includegraphics[width=7.0cm]{xx.eps} \par}
% \caption[Acoustic response for three different cementation schemes.]
% {\small\label{Fig12}
% }
% \end{figure}

Cements can also accumulate  as a {\it coating cement} on grain surfaces, or at pore bodies far from contacts. 
The former, can be the case when cement precipitates as a solid material in low flux zones (Manmath and Lake,  1995). The latter case, is encountered in some  sands which held together by confining pressure only. 
Following (Dvorkin and Nur, 1996), this kind of cement is termed here as {\it friable} or {\it pore body cement}. 
This geometric character of cementation process can be explored in our simulations by following  details  given in section \ref{coatingcement} for {\it coating cement} and section \ref{friable} for {\it friable cement}. Methods to simulate the case of cement at contacts, treated previously, are discused in section \ref{contactcement}. 

In Figure \ref{Fig12}, the lowest velocity trend is obtained  when solid cement is added preferentially as pore-filling material.  Since this kind of cement accumulates far from contacts,  it contributes marginally to rock stiffness. The main effect of this kind of cement is to reduce porosity.
 As more pore-filling material is added, cement particles start to bridge separated grains. Then,  blocks of cement particles are also compressed during pulse propagation with the net effect of  an increasing  slope of velocity trend as porosity reduces. 

It is to be noticed that if clays were added during settling, we   should expect a different behaviour. In that case, above a critical
 concentration of clays (about $\approx 40\%$), they  become load-bearing and completely surround disconnected  quartz grains. 
In such case, soft clays dominate the acoustics and velocity decrease for porosities below some thereshold value (Hezhu, 1992). 

In the case treated here, Figure \ref{Fig12} shows a monotonic   velocity increase with cement saturation. Here, clays or any other cementing material is added after settling. Therefore, quartz grains remain in contact in the whole porosity range. As a consequence, soft clays may reduce porosity but  sound velocity can only decrease slightly due to inertial effects. 
This result suggests that for low porosity sands with high clay content, high sound velocity should be expected if clays are of diagenetical origin.  Low porosity-velocity can be related to the depositional origin of clays. This may be an important factor to consider when  reconstructing the history of a  reservoir sample.

The velocity trend for coating cement lies in between limiting cases of friable and  contact cement. In this case, a fraction of cement deposits near contacts and contributes to rock stiffness, while a significant volume of cement mainly reduces porosity. Figure \ref{Fig12} shows that for the same porosity, this kind  of cementation leads to higher velocities than those of friable sand 
and lower velocities than those of the contact-cemented sample. 
These results  show that  microstructural details are very important parameters which influence in the velocity-porosity trends of cemented samples.

\begin{figure}[!h]
\begin{center}
\center{\par\centering
\includegraphics[width=8cm]{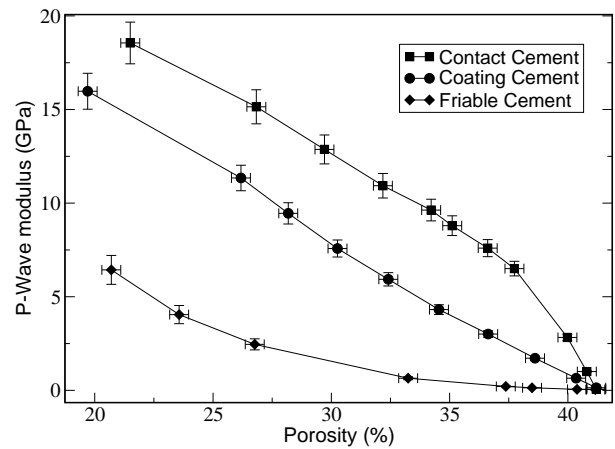} \par}
\caption[P-Wave modulus against porosity for three different cementation schemes.]
{\small\label{Fig13}
In the medium to high cement saturation, contact cement leads to a 
quasilinear trend for P-Wave modulus with porosity in agreement with experimental results (Dvorkin et al., 1994b). Coating cement and friable sands 
shows an upward concavity observed in some naturally occurring sands (Dvorkin and Nur, 1996). 
}
\end{center}
\end{figure}

To further compare our results with available experimental data, we calculated the P-Wave modulus from results in Figure \ref{Fig12}. Computed moduli are shown against porosity in Figure \ref{Fig13}. 

% \begin{figure}[!h]
% \center{\par\centering
% \includegraphics[width=7.0cm]{xx.eps} \par}
% \caption[P-Wave modulus against porosity for three different cementation schemes.]
% {\small\label{Fig13}
% }
% \end{figure}

As expected, results in Figure \ref{Fig13} show different 
trends for P-Wave modulus depending on the amount of cement and cementation scheme. In the case of contact cement, a quasilinear trend is observed for cement saturations above $\approx 10\%$. This result, qualitatively coincides with previous experimental results reported in (Dvorkin et al., 1994b) and references therein. Similar behavior was observed for the case of coating cement. Nevertheless, P-modulus is smaller for the range of cement saturation  studied. Velocity trends obtained for the case of friable cement can be compared with experimental data obtained for some samples with scarce intergranular cement (Dvorkin and Nur, 1996). These results are in good agreement with the varios correlations  between P-modulus and porosity observed in reservoir sands.  

 It must be noted that details of microstructure  are hardly  included in most theoretical approximations. Some   models  treat the case of two a phase composite and/or spherical inclusions, being useless in the range $0\%<S_c<100\%$. For instance, Hashin-Shtrikman model establish bounds  to the maximum and minimum elastic moduli at zero porosity ($S_c=100\%$). Although a generalization of this model would allow us to calculate bounds for the moduli at intermediate porosities (Mavko et al., 1988), in our simulations and in the experiments considered, voids are filled with air with negligible elastic modulus. In this case, lower bound for the moduli is trivial(zero) in all the range of cement saturation considered.  Although Contact Cement Model explicitly includes the microstructure, some the assumptions made in the theory are not supposed to be valid for large amounts of cement.

\section{Conclusions}

Simulation results show that cements modify elastic properties of sands in  different manner, depending on the relative stiffness of  cement and the hosting medium, the amount of cement and its  localisation within pores.

If cement is added preferentially near contacts, it can contribute noticeably to the stiffness of the composite, the effect being  enhanced in uncompacted granular samples. Friable cement leads to relatively low velocity samples even when for moderate cement saturations. Velocity trends show characteristic concavities in each case. 

As confining  pressure increases, the hosting medium for the cement becomes stiffer by compaction and cements have an smaller effect on the sound velocity of the final composite. As a result, velocity-porosity trends differ for   samples that follow different compaction-cementation paths. Although contact cement theory does not account for the confining pressure effects, it can  give a good estimate for the sound velocity in precompacted samples with small amounts of contact cement($S_c<15\%$).

Simulation results presented,  
qualitatively and quantitatively reproduce the available experimental data  where porosity reduces and velocity increases due to cementation. The proposed methods, are suited to take into account the microstructural details of cementation process and their acoustical implications. As these details are hardly included in most effective medium approximations, simulation techniques can be considered an alternative tool to model the acoustics of cemented sands and to get new insights on the underlying physics behind the phenomena.

\section*{Nomenclature}

 \begin{description}

 \item{ 
 $FZI$  Geometric index to characterise microstructure}

 \item{
 $P_s$ Pore shape related value
 }

 \item{
 $S_v$, Pore specific surface
 }
 
 \item{
 $\tau$ Tortuosity
 }

 \item{
 ${\bf F}_c$  Contact force  }
 
 \item{
 ${\bf F}_n$  Hertz Force for a pair of grains in contact }
 
 \item{
 ${\bf F}_s$  Shear force of a grain-to-grain contact }
 
  \item{
 ${\bf \Delta F}_s$ Change in Shear force due to a perturbation}
 
 \item{
 ${\bf \Delta F}_n$ Change in Normal force due to a perturbation }
 
 \item{
 $\Delta{\bf F}_c=\Delta{\bf F}_n+ \Delta{\bf F}_n$
 }

 \item{
 $a_c$        Radius of the contact area   
 }
 
 \item{
 $E_i, E_{f}$ Young modulus of grain $i$ and effective Young modulus for a contact  }

 \item{
 $\kappa_n$   Hertz normal stiffness of a contact}
 
 \item{
 $\kappa_s$   Shear stiffness of a grain-to-grain contact
 }
 
 \item{
 ${\bf\hat{n}}_{12}$ Unitary vector joining the centre of grains $1,2$.
 }
 
 \item{
 $\hat{s}$ Unitary vector tangential to the contact 
 }

 \item{
 $\nu,\nu_i, \nu_c$  Poisson coefficient of grains, particle $i$ and cement
 }

 \item{
 $R, R_f$ Grain radius and effective radius for Hertzian contacts
 }
 
 \item{ 
 ${\bf r}_i$  Position vector of grain $i$
 }

 \item{
 $\mu$ Poisson coefficient of dry friction between grains
 }
 
 \item{
 $e_n$ Restitution coefficient for normal interaction
 }

 \item{
 $\gamma_n, \gamma _s$ Damping constants for normal and tangential directions  to the contact.
 }

 \item{
 $c_w$ Width of cells that discretise the sample
 }

 \item{
 $\rho _c, \rho_g$ Volumetric mass density of cement and grains
 }

 \item{
 $\xi$    Overlapping between spheres in contact  
 }
 
 \item{
 $\xi_0$   Overlapping between spheres in mechanical equilibrium  
 }

 \item{
 $\zeta$  Relative displacement of contacting particles in direction tangential to the contact
 }
 
 \item{
 $\zeta_0$  Relative displacement of contacting particles in direction tangential to the contact when particles are in mechanical equilibrium
 }

\item{
$G$, $G_c$   Shear modulus of material composing grains and cement
}
 
\item{
$\lambda_c$ Lam\'e constant of cementing material
}
 
\item{
$C_n$  Normal stiffness for cement particles interaction   
}
 
 \item{
$C$			Elastic constant for the interaction between cement particles 
}

\item{
${\bf x}_{12}$ Vector joining particles in the undisturbed cement lattice
 }
 
\item{
${\bf u}_{12}$ Relative displacement of particles from their positions on the undisturbed lattice. 
}

\item{
 $m_c$ Mass of a cement particle
}

\item{
$V_{pc}, V_{sc}$ P-wave and S-Wave velocities of cementing material
}

\item{
$V_{p}$ P-wave velocity of the grains and cement composite
}

 \item{
 $P_c$ Confining pressure
 }
 
 \item{
 $\phi$ Porosity of the cemented sample.
 }
 
 \item{
 $\phi_o$ Porosity before cementation.
 }

\item{ 
$S_c$.  Cement saturation
}

\item{
$K_f, G_f,M_c$  Bulk, Shear and Compressional modulus of cement
}

\item{
$\Omega$ Computational cost
}

\item{
$t_c$ Characteristic time for the interaction among particles
}

\end{description}

\section{Acknowledgments}
This work was supported by the IVIC Rocks
project and by FONACIT through grant S1-2001000910.
% 
% We would like to thank the Venezuelan Institute of Scientific Research  
% for sharing the facilities to make this research. To the free software 
% community for the great development tools and imagery related 
% packages made available worldwide.

\appendix

\section{Effective medium approximation for cemented sands}
\label{emt}
According to the theoretical model proposed by Dvorkin et al. 
(1994a); Dvorkin et al. (1994b) and Dvorkin and Nur(1996), the 
initial porosity 
$\phi_o$ of an uncemented sample is decreased to $\phi$ by the 
addition of cementing material. Once a given volume of cement is 
added, the effective bulk $K_f$ and shear $G_f$ moduli of the composite  
is calculated according with Eq. \ref{Keff}.

\begin{eqnarray}\label{Keff}
  \begin{array}{l}
 K_f=\frac{1}{6}N(1-\phi_o)M_c \hat{S_n}\\ 
 \\ 
 G_f=\frac{3}{5}K_f+\frac{3}{20}N(1-\phi_o)G_c\hat{S_t}\\
 \\
 M_c={V_{pc}}^2\rho_c\\
 \\
 G_c={V_{sc}}^2\rho_c,
  \end{array}
\end{eqnarray}
where $\rho_c, V_{pc}, V_{sc}$ are mass density, P-wave and 
S-wave  velocities of cementing material and $N$ is the average
number of contacts per grain. The normal and tangential stiffness
of the contacts in the cemented sample, denoted as $\hat{S}_n$
and $\hat{S}_t$ respectively, are given as function of porosity
change due to cementation in Eq. \ref{Stiffness} below.
\begin{equation}\label{Stiffness}
  \begin{array}{l}
	\hat{S}_n=A_n\alpha^2 + B_n\alpha + C_n\\
	\hat{S}_t=A_s\alpha^2 + B_s\alpha + C_s
  \end{array}
\end{equation}
where: 
\begin{equation}\label{beta}
\begin{array}{l}
 
 \alpha= \left \{ \begin{array}{ll} 
         
\left ( \frac{2}{3}\frac{\phi_0-\phi}{1-\phi_0}\right)^{1/2}
         &\textrm{ for cement in layers}\\
        
2\left(\frac{1}{3N}\frac{\phi_0-\phi}{1-\phi_0}\right)^{1/4}
         &\textrm{ for cement at contacts}\\
  \end{array}
  \right \}  
\end{array}
\end{equation}

The coefficients of the $\alpha$ powers in Eq.
\ref{Stiffness} are given in formulae \ref{formulas}:
\begin{equation}\label{formulas}
 \begin{array}{l}
  A_n=-0.024153 {\Lambda}_n^{-1.3646}\\
  B_n=0.20405 \Lambda _n^{-0.89008}\\
  C_n=0.00024649 \Lambda _n^{-1.9864}\\
  \\
A_s=-10^{-2}(2.26\nu^2+2.07\nu+2.3)\Lambda_s^{\omega_1}  \\
B_s=(0.0573\nu^2+0.0937\nu+0.202)\Lambda_s^{\omega_2}    \\
C_s=-10^{-4}(9.654\nu^2+4.945\nu+3.1)\Lambda_s^{\omega_3}\\
\\
\textrm{where}\\
\\
\omega_1={(0.08167\nu^2+0.4011\nu-1.8186)}\\
\omega_2={(0.027\nu^2+0.0529\nu-0.8765)}\\
\omega_3={(0.079\nu^2+0.1754\nu-1.342)}\\

\Lambda _n=\frac{2G_c}{\pi G}\frac{(1-\nu)(1-\nu_c)}{1-2\nu_c}\\
\Lambda_s=\frac{G_c}{\pi G},
 \end{array}
\end{equation}
where $\nu,\nu_c$ are respectively the grains and cement Poisson
ratio, $G$ is the shear modulus of grains and the rest of symbols
have the same denotation than in previous equation.

The theoretical predictions for the P-Wave velocity are then
obtained for the cemented sample by substituting  elastic
moduli given in Eq. \ref{Keff} in linear elasticity
Eq. \ref{vple}

\begin{equation}\label{vple}
V_p=\left( \frac{K_f + 4/3 G_f}{\rho _f} \right )^{1/2},
\end{equation}
where $V_p$ is the sound velocity of the composite
system of grains and cement and $\rho _f$ the effective 
mass density. In our simulations, we have used Eq. \ref{vple} to 
fit the obtained data.

\section{The effect of resolution in the computational cost}
\label{resolutioncost}

Resolution $\Omega$ affects computational cost in three different
ways. First, cells width $c_w \approx R/\Omega$, being $R$
the average grain radii. As $\Omega$ increases, $c_w$ reduces
and the number of cells to discretise the sample increases as 
$\approx \Omega ^3$ in three dimensions. The larger the
number of cells the larger the memory needed to store the
information related with each cell. Additionally, as the
number of cement particles raise, the number of operations
to compute forces, velocities and actualise positions
increases proportionally. 

On the other hand, for smaller cells widths $c_w$, the 
mass of cement particles goes as $m_c \approx \Omega ^{-3}$ 
and elastic stiffness constants goes as $\kappa \approx
\Omega ^{-1}$. The characteristic integration step $t_c$ is
determined by the characteristic time related to interactions
$t_c \approx (m_c / \kappa)^{1/2}\approx \Omega$.
Therefore, as resolution $\Omega$ increases, the number of
integration steps to simulate pulse propagation in the system 
during a unitary time rises linearly. As both the number of
cement particles and integration steps increase, the
computational cost is higher.

\end{document}